\documentstyle[sprocl]{article}

\input{psfig.sty}

\def\Journal#1#2#3#4{{#1} {\bf #2}, #3 (#4)}

\def\PLB{{\em Phys. Lett.}  B}

\def\PRD{{\em Phys. Rev.} D}

\def\be{\begin{equation}}
\def\ee{\end{equation}}
\def\bea{\begin{eqnarray}}
\def\eea{\end{eqnarray}}

\begin{document}

\title{Spectroscopy ``windows'' of quark-antiquark mesons and
                     glueballs with effective Regge trajectories.  }

\author{Martina Brisudova}

\address{Physics Dept., University of Florida, Gainesville, \\ FL 32611, USA} 

\author{Leonid Burakovsky, Terrance Goldman}

\address{Theoretical Division, T-16, Los Alamos National Laboratory, Los 
Alamos,\\ NM 87545, USA }


\maketitle\abstracts{ Regge trajectories of quark-antiquark mesons can be well 
approximated for phenomenology purposes by a specific nonlinear form, 
reflecting that the flux tubes cannot be arbitrarily large, but break due to 
the effect of pair-production. If confirmed, this would imply that there is 
only a finite number of bound states on each trajectory, and consequently, an 
existence of ``spectroscopy windows'' for each flavor. Here we present our 
results for these windows.}

\section{}
It is a paradigm in QCD that Regge trajectories (roughly speaking, functions 
that relate angular momentum of a state to its mass squared) of mesons are 
linear. This 
picture arises in the Veneziano model for the amplitudes, string theory of 
hadrons and also, at least asymptotically, from potential models with a linear 
confining potential.

On the other hand,  Regge trajectories extracted from data are {\it 
nonlinear.} 
In addition to being 
disfavored by 
 experiments, linear trajectories also 
lead to problems in theory. (For details on both experimental and theoretical 
aspects, and references, see \cite{us}.)

All the theoretical concepts that lead to linear Regge trajectories in QCD 
overlook one feature of QCD - the production  of color singlet pairs. Pair 
production is only virtual at short distances, but as the energy of the flux 
tube increases, it is possible (and likely) to create a real pair and the 
breaking of the flux tube is energetically favorable.  
One can  argue that 
 hadronic Regge trajectories cannot rise indefinitely, and they acquire a 
curvature  due to 
pair production which screens the confining QCD potential at large distances 
\cite{us0}. 
Once the nonlinearity of Regge trajectories is an accepted fact, 
the question of what specific form should be used for phenomenology arises.

We have considered a whole class of nonlinear trajectories allowed by dual 
amplitudes with Mandelstam analyticity (DAMA) (\cite{Jenk} and references 
therein).
 DAMA 
allow for 
Regge trajectories of a form
\begin{eqnarray}
\alpha_{j\bar{i}} (t) & = & \alpha _{j\bar{i}}(0)+\gamma \Big[ 
T_{j\bar{i}}^\nu -(T_{j\bar{i}}-t)^\nu \Big]   
\label{trajectory}
\end{eqnarray} 
where $\nu$ is a constant restricted to $0 \leq \nu \leq {1\over{2}}$; 
$\gamma $ is a universal constant; $T_{j\bar{i}}$ is a trajectory threshold, 
$\alpha_{j\bar{i}}(0)$ is its intercept, and $i, j$ refer to flavor. 
For any $\nu \neq 0$ in this interval, the trajectory becomes purely imaginary 
at the trajectory threshold (in other words, its real part terminates); for $\nu 
=0$ it develops a constant imaginary part. This means that any $\nu \neq 0$ 
trajectory supports only a finite number of bound states, with their  value of 
angular momentum limited by $\alpha (T)$. Beyond the threshold,  there are  
continuum states. This picture seems to have captured the essence of the effect 
of pair production.

The largest deviations of the true trajectory from any of these forms can be 
expected for the states near the threshold. 
The sensitivity of the parameters on the specific form (i.e. the value of $\nu$) 
is 
also 
interesting. We found that the least sensitive is the threshold (up to few 
percent), and the most sensitive is the resultant maximum angular momentum (i.e. 
$\alpha (T)$ for any $\nu \neq 0$; for $\nu =0$ it is infinite)
\cite{us}. This means, in our opinion,  that while the maximum angular momentum 
for a given flavor cannot be predicted, the values of thresholds extracted 
from data can be taken seriously, regardless of what DAMA form is assumed.
We have argued that both limiting cases ($\nu =1/2$ and $\nu=0$) can be 
expected to work comparably well for lowest lying states (and thus, any $\nu$ in 
between), but the $\nu=1/2$, 
so-called square-root, form  is likely to be more realistic.
Therefore, we use the square-root form for phenomenological purposes.

Assuming that Regge trajectories are of the  form (\ref{trajectory}) with $\nu 
=1/2$, we determine thresholds and intercepts of trajectories by using various 
experimental information. Typically, we use masses of a few known lowest lying
states, and in the case of the $\rho$ trajectory we also use the value of the 
intercept (which is known and well-established) found from exchange processes.
The value of $\gamma$ (the universal asymptotic slope) is fit to the $\rho$ 
trajectory, and then taken as universal for all other trajectories.

The approach has more predictive power than one would naively expect. The number 
of parameters for any two parity related 5-flavor multiplets\footnote{An example 
of a pair of parity partners trajectories is the $\rho$ and $a_2$ trajectories; 
the 5-flavor multiplet containing the $\rho$ trajectory includes $K^*$, $\phi$, 
$D^*$, $D_s^*$, $J/\psi$, $B^*$, $B_s^*$, $B_c^*$ and $\Upsilon$ trajectories.}  
 (i.e. 20 trajectories in all) is only 12, in contrast to 40.
 
\begin{figure}[t]
\psfig{figure=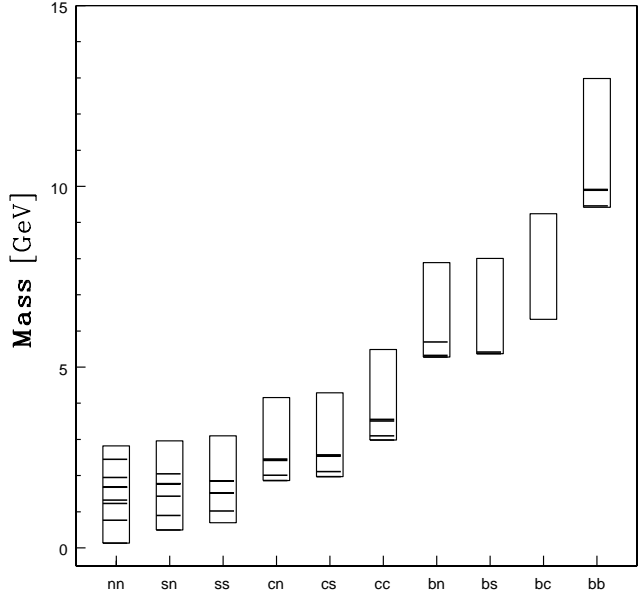,height=3.5in}
\caption{ Mass windows for quark-antiquark states of a listed flavor, together 
with known masses of states belonging to the vector-, tensor-, pseudoscalar- and 
axial-vector- trajectories. $n$ refers to light quarks. For example, the $nn$ 
tower contains $I=1$ mesons $\pi$, $\rho$, $a_1$, $a_2$ etc. \label{fig:figure}}
\end{figure}

I would like to concentrate here on just one of the numerous consequences and 
implications. Specifically, if we are right, there exist ``spectroscopy 
windows'' as shown in Figure \ref{fig:figure}. This may simplify identification 
of 
states observed in experiments,  and might be beneficial in searches for exotics.
 
\section*{Acknowledgments}
This research is supported in part by the U.S.D.O.E.   
under contract W-7405-ENG-36, and grant DE-FG02-97ER-41029.

\end{document}